\documentclass[10pt,reprint,aps,pra,twocolumn,notitlepage,nofootinbib,superscriptaddress,noeprint]{revtex4-2}
\usepackage[utf8]{inputenc}
\usepackage[english]{babel}
\usepackage[T1]{fontenc}
\usepackage{graphicx}
\usepackage{amssymb}
\usepackage{bm}
\usepackage{mathtools}
\usepackage{physics}
\usepackage[caption=false]{subfig} 
\usepackage{xcolor}
\usepackage{dsfont}
\usepackage{ulem}
\begin{document}

\preprint{APS/123-QED}

\title{Photon-Pair Advantage for Dissipative Parameter Estimation in Whispering-Gallery-Mode Interferometers}%
\author{Elnaz Darsheshdar}
\thanks{These authors contributed equally to this work.}
\email{darsheshdare@gmail.com}
\altaffiliation{Department of Physics and Astronomy, University of Exeter, Stocker Road, Exeter, EX4 4QD, UK} 
\altaffiliation{Institute for Physical Research, Ashtarak-2, 0203, Armenia}
\author{Callum Jones}
\thanks{These authors contributed equally to this work.}
\email{c.h.jones@exeter.ac.uk}
\altaffiliation{Living Systems Institute, University of Exeter, Stocker Road, Exeter, EX4 4QD, UK; Department of Physics and Astronomy, University of Exeter, Stocker Road, Exeter, EX4 4QD, UK}
\author{Seyed Mostafa Moniri}
\altaffiliation{Basic Sciences Group, Golpayegan College of Engineering, Isfahan University of Technology, Golpayegan 87717-67498, Iran}
\author{Antonio Vidiella-Barranco}
\altaffiliation{Gleb Wataghin Institute of Physics - University of Campinas, 13083-859 Campinas, SP, Brazil}
\author{Jolly Xavier}
\altaffiliation{Living Systems Institute, University of Exeter, Stocker Road, Exeter, EX4 4QD, UK; SeNSE, Indian Institute of Technology Delhi, Hauz Khas, New Delhi, India}%
\author{Frank Vollmer}
\altaffiliation{Living Systems Institute, University of Exeter, Stocker Road, Exeter, EX4 4QD, UK}

\date{\today}

\begin{abstract}
We study the estimation of detuning and intrinsic loss in a lossy whispering-gallery-mode (WGM) resonator located in a balanced Mach--Zehnder interferometer. We use indistinguishable photon pairs and coherent-state probes.
We show that using indistinguishable photon pairs can provide higher Quantum Fisher Information (QFI) of dissipative parameters, such as intrinsic loss, at resonance away from critical coupling.
Taking into account all photon-loss outcomes, we obtain analytical QFI expressions and compare photon-pair and coherent-state probe results when they have the same mean photon number.
For estimating the round-trip amplitude transmission at resonance and away from critical coupling, the photon-pair probe gives a higher QFI than the coherent-state probe, and photon-number-resolved detection saturates this QFI. This advantage is due to the photon-loss statistics and it is still accessible for a typical overcoupled condition after averaging over fluctuations of the unresolved detuning. 
We identify the limits of this advantage: phase doubling does not improve resonant detuning estimation, and coincidence-only detection is inefficient close to a zero transmission. These results identify the cases in which photon pairs can be useful for absorption and pump--probe WGM sensing.

\end{abstract}

\maketitle

\section{Introduction}

Whispering-gallery-mode (WGM) resonators are high-$Q$ optical cavities where light circulates around a curved dielectric interface.  They have various shapes, such as rings, toroids, spheres, and bottles, and are widely used in sensing of chemical and biological ~systems\cite{Jiang_Qavi2020,Yu_Humar2021,Loyez_Adolphson2023}.

A perturbation of a WGM can be detected through a dispersive response, such as a resonance-frequency shift or an optical phase change, or through a dissipative response, such as a change in linewidth, transmission, intrinsic loss, or absorption. Despite the fact that these responses may occur simultaneously in a physical system, identifying each contribution is important as they need not carry the same amount of metrological information.

Several methods have been investigated to improve the sensing of the WGM. Plasmonic near-field enhancement affects the local light--matter interaction and can enable the detection of single molecules and enzyme turnover events~\cite{Santiago-Cordoba2011,He_Ozdemir2011,Baaske_Vollmer2016,Vincent_Subramanian2020,Subramanian_Jones2021}. 
More recently, plasmon-enhanced WGM microlasers have made it possible to detect single ions without adding any labels or markers, because the ion itself changes the local optical field and therefore the WGM signal~\cite{VartabiKashanian2026}.

There are other methods to engineer the optical response itself, for instance operating near exceptional points or critical-coupling transmission zeros~\cite{Chen_Ozdemir2017_EP,Wiersig2014_EP,Cai_Painter2000_CriticalCoupling}. Notably, the enhancement discussed here is obtained in a fully passive resonator, without optical gain or operation near a lasing threshold. 
However, a narrow spectral response, a large phase gradient, or a very small transmission amplitude does not necessarily lead to better precision of estimation, because the available information depends on the statistics of the output light, noise, measurement strategy, and resources.

Quantum sensing investigates whether nonclassical probes offer a genuine improvement over classical ones ~\cite{Degen2017QuantumSensing,Giovannetti2004QuantumEnhanced,Pezze2018QuantumMetrology,Darsheshdar2024,Albarelli2023}. The quantum Fisher information (QFI) sets the ultimate precision on estimation that is allowed by quantum mechanics, while the classical Fisher information (CFI) quantifies the information accessible through a specified measurement~\cite{Moniri2026Magnetometry}. A meaningful comparison between quantum and classical probes needs both an equal-resource constraint and a clear treatment of optical loss. In this work we use these methods to evaluate the sensing enhancement using indistinguishable photon pair probes in a WGM resonator coupled to one arm of a Mach--Zehnder interferometer (MZI).

In an MZI, applying one indistinguishable photon into each input arm leads to a two-photon N00N state after the first beam splitter because of Hong--Ou--Mandel interference. A relative phase $\phi$ between the two arms is then accumulated as $2\phi$, producing fringes with twice the classical oscillation frequency. This phase super-resolution has been demonstrated with two- and higher-photon N00N states~\cite{Nagata_Okamoto2007,Crespi_Lobino2012,Slussarenko_Weston2017}. In fact, a doubled phase response is not enough to have a quantum metrological advantage, especially in a lossy interferometer where the probability as well as coherence of the transmitted multiphoton component are reduced.

Loss estimation has a different metrological problem. Fixed-photon-number probes can practically be better than coherent states in the estimation of bosonic-channel transmission since individual photon-loss events provide additional information about the unknown attenuation~\cite{Monras_Paris2007,Adesso_DellAnno2009}.

In the cases where absorption and phase are both encoded~\cite{Birchall_Allen2020}, nonclassical advantages have been studied. 
In contrast, a coherent-state light in a ring resonator was shown to outperform any single-pass quantum probe since the resonance allows the light-matter interaction to be stronger, even when the same mean number of photons is used~\cite{Belsley_Allen2022}. Further, at critical coupling and on resonance, bright single-mode squeezed states were shown to offer no advantage over coherent-state probes for estimating refractive index or absorption in an all-pass ring resonator~\cite{Belsley_Allen2022}.
 These results show that nonclassical light does not always have a better sensing performance: the advantage depends on both the optical architecture of the sensing and how the parameter is encoded in the scattered light.

The main question is how the advantage of fixed-photon-number states for loss estimation compares with the resonantly enhanced phase response of a WGM--MZI, especially near zero transmission and when only limited photon-counting measurements are available. We address this question in a system of a WGM resonator coupled to one arm of an MZI. The resonator introduces a complex transmission amplitude of $t=\sqrt{T}e^{i\phi}$, thus an unknown parameter may be encoded in the transmission magnitude $T$, the phase $\phi$, or even both of them.

We first formulate the lossy WGM channel in the case of the coherent-state, and then derive a loss-consistent photon-pair output state that can have the zero-, one-, and two-photon loss sectors. After tracing out the environmental modes, we compare this state with a coherent-state probe having the same mean photon number. We separate the total QFI into contributions related to changes in transmission and phase. For estimation of the WGM detuning, the doubled two-photon phase and the resulting spectral response do not, by themselves, yield an advantage over the coherent-state benchmark. 
On the other hand, when the parameter is encoded in the intrinsic loss, the fixed-photon-number structure of the photon-pair probe can have a genuine enhancement due to the statistics of the different photon-loss sectors.

We use this point for estimating the WGM round-trip amplitude transmission $\alpha$ and relate this to intrinsic loss, absorption, and absorbance. In addition we determine how much of the available information can be accessed using coincidence-only and photon-number-resolved detection, and examine the robustness of the resulting advantage to unresolved shot-to-shot detuning fluctuations. This specifies the conditions under which photon pairs have a practically accessible advantage for dissipative WGM sensing and motivates applications to absorption-based and pump--probe measurements.

\begin{figure}
    \centering
    \includegraphics[width=\columnwidth]{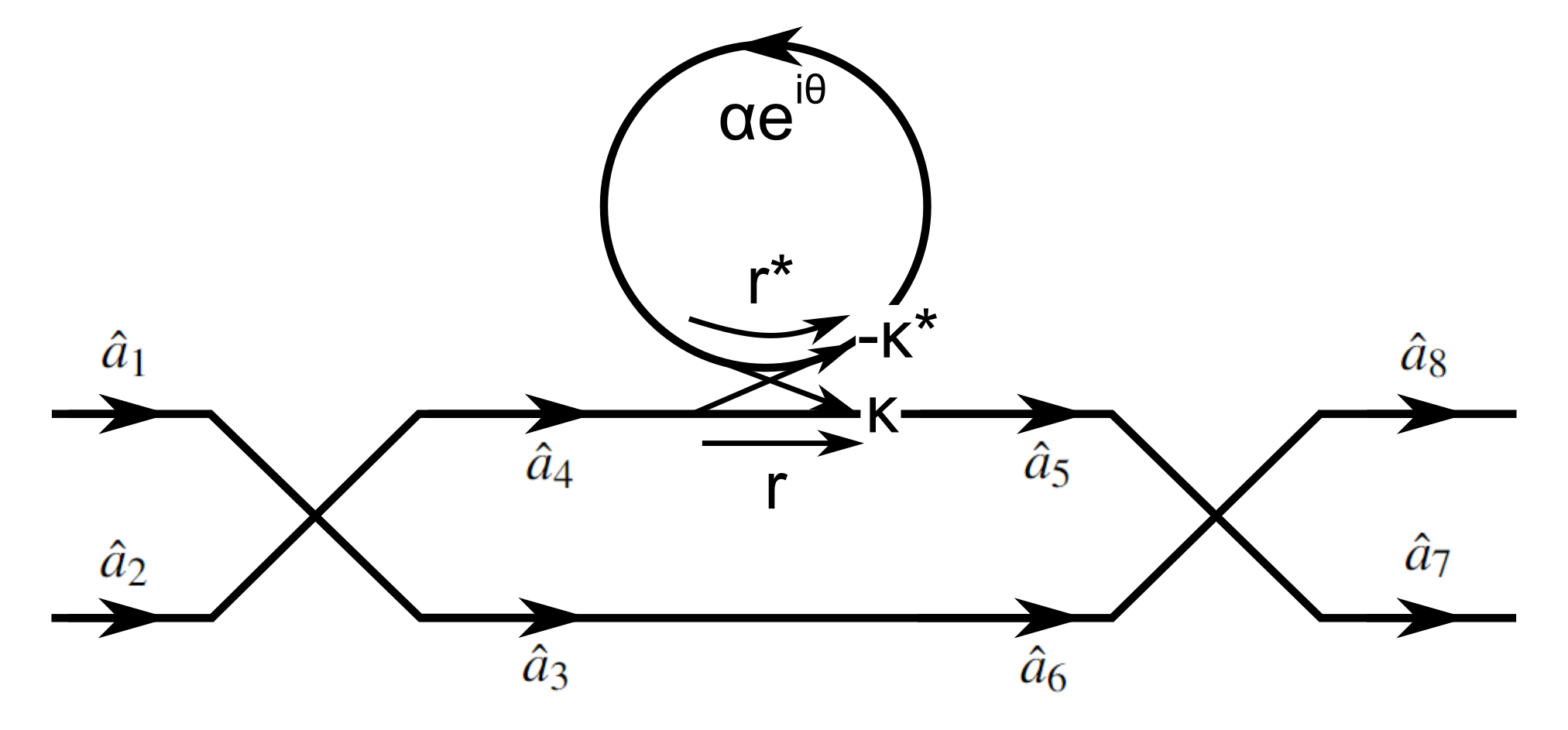}
    \caption{Schematic of the WGM resonator coupled to one arm of the Mach--Zehnder interferometer (MZI). The WGM introduces a detuning-dependent complex transmission amplitude $t(\delta)$ in one arm, while the other arm acts as a reference path.}
    \label{fig:WGM_MZI_schematic}
\end{figure}

\section{Quantum Optical Model of the WGM--MZI}
\label{sec:quantum_model}

\subsection{WGM transmission and loss}

Figure~\ref{fig:WGM_MZI_schematic} illustrates a schematic sensing geometry and labels of each mode. We use the operator-valued phasor-addition (OVPA) method in Ref.~\cite{Alsing_Hach2017} for describing the coupling between a travelling mode of the waveguide and a lossy WGM resonator. The directional coupler is characterised by the cross-coupling amplitude $\kappa$ and the self-coupling amplitude $r$. We choose the coupler phase convention such that $r$ is real and non-negative, with $|\kappa|^{2}+r^{2}=1$.

The WGM round-trip field transmission is denoted by $\alpha$, with $0<\alpha\leq1$. For a uniform intrinsic intensity-attenuation coefficient $\Gamma(\omega)$ and a round-trip length $L=2\pi R$,
\begin{equation}
\alpha(\omega)=\exp\!\left[-\frac{\Gamma(\omega)L}{2}\right],
\end{equation}
so that $\alpha^{2}$ is the round-trip intensity transmission. We consider $\alpha$ to be approximately constant over the narrow spectral range considered here.

We define the angular-frequency detuning as $\delta=\omega-\omega_{0}$, where $\omega_{0}$ is the WGM resonance frequency. We neglect the material dispersion over the resonance linewidth, thus the detuning-dependent round-trip phase is
\begin{equation}
\theta_{\mathrm{WGM}}(\delta)=\frac{nL}{c}\,\delta
=\frac{2\pi Rn}{c}\,\delta .
\label{eq:theta_wgm}
\end{equation}

The input--output relation for the  mode of the waveguide which is coupled to the WGM is
\begin{equation}
\hat a_{\mathrm{out}}(\omega)=
t(\delta)\hat a_{\mathrm{in}}(\omega)+
\hat F(\omega),
\label{eq:WGM_input_output}
\end{equation}
where $\hat F(\omega)$ is the noise operator related to loss to the environment. It also preserves the bosonic commutation relation for the output field.

The complex WGM transmission amplitude is
\begin{equation}
t(\delta,\alpha,r)=
\frac{r-\alpha e^{i\theta_{\mathrm{WGM}}(\delta)}}{1-r\alpha e^{i\theta_{\mathrm{WGM}}(\delta)}} .
\label{eq:WGM_transmission}
\end{equation}
It is useful to separate the magnitude and phase responses of the complex WGM transmission amplitude:
\begin{equation}
t(\delta,\alpha,r)=\sqrt{T(\delta,\alpha,r)}e^{i\phi(\delta,\alpha,r)}.
\label{eq:t_T_phi}
\end{equation}
Here $T=|t|^{2}$ is the intensity transmission and $\phi=\arg(t)$ is the phase response. The resonator is undercoupled in $r>\alpha$ , critically coupled in $r=\alpha$, and overcoupled for $r<\alpha$ values. At critical coupling and zero detuning we have $t(0)=0$.

The environmental input modes are assumed to be in the vacuum state, as their thermal occupation is negligible at optical frequencies. We also define the corresponding loss amplitude as $\ell(\delta)=\sqrt{1-|t(\delta)|^{2}}=\sqrt{1-T(\delta)}$. The explicit OVPA representation of $\hat F(\omega)$ and its reduction to a normalized effective environmental mode are given in Appendix~\ref{app:photon_pair_derivation}.

\subsection{Coherent-state benchmark}

For the classical light case, we consider a coherent state with a complex amplitude $\beta$.  It enters in port 1 of the MZI, while input port 2 is in the vacuum state:
\begin{equation}
|\psi_{\mathrm{in}}^{\mathrm{coh}}\rangle=|\beta\rangle_{1}\otimes|0\rangle_{2},
\label{eq:coherentinput}
\end{equation}
and $|\beta|^{2}$ is the mean photon number. Linear optical transformations and vacuum loss map coherent states to coherent states. Therefore, after tracing over the undetectable environment, the detected output is a product of coherent states. With the beam-splitter convention used in Fig.~\ref{fig:WGM_MZI_schematic}, the output coherent amplitudes in modes 7 and 8 are $i[1+t(\delta)]\beta/2$ and $[1-t(\delta)]\beta/2$, respectively.

The corresponding mean output photon numbers, normalized by the
incident mean photon number $|\beta|^{2}$, can be written as
\begin{equation}
P_{\pm}^{\mathrm{coh}}(\delta)
\equiv
\frac{\langle\hat n_{7,8}\rangle}{|\beta|^{2}}=\frac{1}{4}\left|1\pm t(\delta)\right|^{2},
\label{eq:class_WGM_MZI}
\end{equation}
where the plus and minus signs correspond to output modes 7 and 8,
respectively.

\subsection{Photon-pair probe and physical loss model}

In the case of nonclassical light probe, one photon enters each input port of the first beam splitter:
\begin{equation}
|\psi_{\mathrm{in}}^{\mathrm{pair}}\rangle=
\hat a_{1}^{\dagger}\hat a_{2}^{\dagger}
|0,0\rangle_{1,2}\otimes|0\rangle_{\mathrm{env}} .
\label{eq:input_state}
\end{equation}
Hong--Ou--Mandel interference at the balanced first beam splitter produces the two-photon N00N state
\begin{equation}
|\psi_{4,3}\rangle=
\frac{i}{\sqrt{2}}\left(|2\rangle_{4}|0\rangle_{3}+
|0\rangle_{4}|2\rangle_{3}\right)
\otimes|0\rangle_{\mathrm{env}},
\end{equation}
where mode 4 is coupled to the WGM and mode 3 is the reference arm.

Applying the unitary loss dilation and the second balanced beam splitter, as derived in Appendix~\ref{app:photon_pair_derivation}, gives the joint state of optical--environmental in the output as  
\begin{widetext}
\begin{equation}
\begin{aligned}
|\psi_{\mathrm{out}}\rangle=&
\frac{i\sqrt{2}}{4}\left[t^{2}(\delta)-1\right]
|2,0\rangle_{7,8}\otimes|0\rangle_{\mathrm{env}}
+\frac{i\sqrt{2}}{4}\left[1-t^{2}(\delta)\right]
|0,2\rangle_{7,8}\otimes|0\rangle_{\mathrm{env}}\\
&+\frac{1}{2}\left[1+t^{2}(\delta)\right]
|1,1\rangle_{7,8}\otimes|0\rangle_{\mathrm{env}}
-\frac{i\,t(\delta)\ell(\delta)}{\sqrt{2}}
|1,0\rangle_{7,8}\otimes|1\rangle_{\mathrm{env}}\\
&-\frac{t(\delta)\ell(\delta)}{\sqrt{2}}
|0,1\rangle_{7,8}\otimes|1\rangle_{\mathrm{env}}
+\frac{i\ell^{2}(\delta)}{\sqrt{2}}
|0,0\rangle_{7,8}\otimes|2\rangle_{\mathrm{env}} .
\end{aligned}
\label{eq:two_photon_output_state}
\end{equation}
\end{widetext}

The normalized environmental Fock states are defined by
\begin{equation}
|1\rangle_{\mathrm{env}}=
\hat b_{\mathrm{out}}^{\dagger}|0\rangle_{\mathrm{env}},\qquad
|2\rangle_{\mathrm{env}}=
\frac{(\hat b_{\mathrm{out}}^{\dagger})^{2}}{\sqrt{2}}
|0\rangle_{\mathrm{env}}.
\end{equation}
The surviving two-photon amplitude in the WGM arm is $t^{2}=T e^{2i\phi}$. It therefore accumulates twice the single-photon phase, but its magnitude is simultaneously reduced by the intensity transmission $T$. The coincidence probability is
\begin{equation}
P_{11}(\delta)=\frac{1}{4}\left|1+t^{2}(\delta)\right|^{2}.
\label{eq:P11}
\end{equation}

\subsection{Reduced detected state and resource normalization}
\label{sec:reduced_detected_state}
 Writing the joint state as
\begin{equation}
|\psi_{\mathrm{out}}\rangle=\sum_{n=0}^{2}
|\chi_{n}\rangle_{7,8}\otimes|n\rangle_{\mathrm{env}},
\end{equation}
where $|\chi_{n}\rangle$ are the unnormalized optical components in Eq.~\eqref{eq:two_photon_output_state}, the detected photon-pair state is
\begin{equation}
\rho_{\mathrm{pair}}=
\operatorname{Tr}_{\mathrm{env}}\left(|\psi_{\mathrm{out}}\rangle\langle\psi_{\mathrm{out}}|\right)
=\sum_{n=0}^{2}|\chi_{n}\rangle\langle\chi_{n}|.
\end{equation}
The environmental modes are not detectable. The probabilities for losing zero, one, and two photons are therefore
\begin{equation}
p_{0}=\frac{1+T^{2}}{2},\quad
p_{1}=T(1-T),\quad
p_{2}=\frac{(1-T)^{2}}{2},
\end{equation}
with $p_{0}+p_{1}+p_{2}=1$. This block structure is useful since a parameter that is encoded in the WGM loss can change both the conditional optical states and the probabilities of the different photon-numbers.

The comparisons between the results of the coherent-state and photon-pair are performed at the same mean incident photon number. Thus the mean-photon-number of the coherent-state benchmark is chosen as $|\beta|^{2}=2$. The mean photon number incident on the WGM-coupled arm, after the first beam splitter,  is one for both probes.

\begin{figure}
    \centering
    \includegraphics[width=\columnwidth]{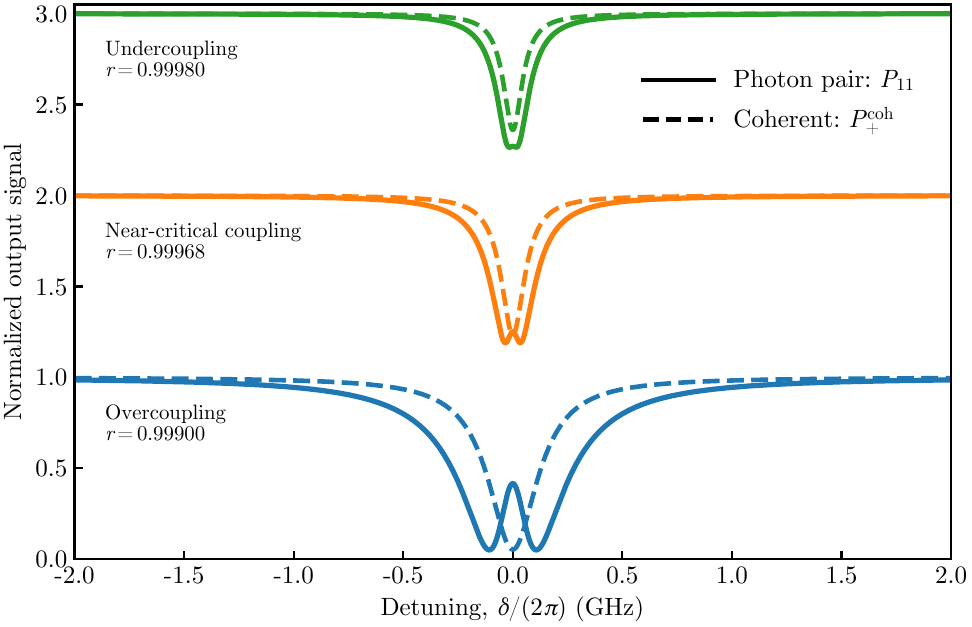}
    \caption{Spectral responses of the WGM-coupled MZI for an indistinguishable photon-pair input and a coherent-state input. Solid lines show the coincidence probability $P_{11}$, while dashed lines show the normalized coherent output intensity $P_{+}^{\mathrm{coh}}$. The round-trip amplitude transmission is fixed at $\alpha=0.9997$, and the coupling coefficient $r$ is varied across the undercoupled, near-critical, and overcoupled regimes. The curves are vertically offset for clarity.}
    \label{fig:wgm_transmission_coinc_a0p9997_varying_r}
\end{figure}

Figure~\ref{fig:wgm_transmission_coinc_a0p9997_varying_r} shows that the two-photon coincidence signal can develop a double-dip structure under a range of coupling conditions. This structure originates from the interference term $t^{2}(\delta)$ and does not require coupling between clockwise and counter-clockwise WGMs, unlike conventional scattering-induced mode splitting~\cite{Kim_Ozdemir2010,Subramanian_Vincent2020}. It is a signature of the two-photon spectral-phase response, but it does not lead to a metrological advantage. The latter must be specified from the QFI as well as the CFI of experimentally accessible measurements.

\section{Quantum Estimation Framework}
\label{sec:QFI}

\subsection{Quantum and classical Fisher information}

We consider the estimation of a physical parameter $\vartheta$  that is encoded in the complex WGM transmission amplitude. The environmental modes are not detectable, thus the detected state is the reduced optical state 
\begin{equation}
\rho_{\vartheta}=\operatorname{Tr}_{\mathrm{env}}\left[|\psi_{\mathrm{out}}(\vartheta)\rangle\langle\psi_{\mathrm{out}}(\vartheta)|\right].
\label{eq:reduced_detected_state}
\end{equation}
In the photon-pair case, this state is the block-diagonal state discussed in Sec.~\ref{sec:quantum_model}; and for the coherent state, it is the two-mode coherent output state defined in the preceding section.

For a measurement with outcomes $y$, if  $p_y(\vartheta)=\operatorname{Tr}[\rho_{\vartheta}\Pi_y]$ denotes the corresponding outcome probabilities, the corresponding classical Fisher information (CFI) is
\begin{equation}
F_C(\vartheta|\{\Pi_y\})=\sum_{y:p_y>0} \frac{\left[\partial_{\vartheta}p_y(\vartheta)\right]^2} {p_y(\vartheta)}.
\label{eq:cfidefinition}
\end{equation}
In cases with vanishing probabilities the CFI will be evaluated using the appropriate limiting procedure.

In the case of a locally unbiased estimator $\widehat{\vartheta}$  which is built from $N_{\mathrm{rep}}$ independent repetitions, the classical and quantum Cram\'er--Rao bounds satisfy
\begin{equation}
\operatorname{Var}(\widehat{\vartheta}) \geq \frac{1}{N_{\mathrm{rep}}F_C(\vartheta|\{\Pi_y\})} \geq \frac{1}{N_{\mathrm{rep}}F_Q(\vartheta)}.
\label{qfi}
\end{equation}
Here, the quantum Fisher information (QFI) is 
\begin{equation}
F_Q(\vartheta)=\operatorname{Tr}\!\left[\rho_{\vartheta}L_{\vartheta}^{2}\right],\qquad
\partial_{\vartheta}\rho_{\vartheta}=\frac{1}{2}\left(L_{\vartheta}\rho_{\vartheta} + \rho_{\vartheta}L_{\vartheta}\right),
\label{eq:lyapunovsld}
\end{equation}
where $L_{\vartheta}$ is the symmetric logarithmic derivative (SLD). Using the spectral decomposition of $\rho_{\vartheta}=\sum_n\lambda_n|e_n\rangle\langle e_n|$, the equivalent expression 
\begin{equation}
F_Q(\vartheta)=2\sum_{\substack{m,n\\\lambda_m+\lambda_n>0}}\frac{\left|
\langle e_m|\partial_{\vartheta}\rho_{\vartheta}|e_n\rangle\right|^2} {\lambda_m+\lambda_n}
\label{eq:qfi_spectral}
\end{equation}
is used for the numerical calculations.

The second MZI beam splitter is independent of $\vartheta$, thus it does not change the QFI. However, it determines how the available information is distributed among the photon-counting outcomes at output modes 7 and 8. Thus we calculate the QFI from the complete reduced state and the CFI for specified output measurements.

For coincidence-only detection, the two outcomes are a coincidence event, with probability $P_{11}(\vartheta)$, and its complement. The corresponding CFI is
\begin{equation}
F_C^{\mathrm{coinc}}(\vartheta)=\frac{\left[\partial_{\vartheta}P_{11}(\vartheta)\right]^2}{P_{11}(\vartheta)\left[1-P_{11}(\vartheta)\right]}.
\label{eq:cfi_coinc_general}
\end{equation}
On the other hand photon-number-resolved (PNR) detection distinguishes all output occupations $(n_7,n_8)$. Having 
\begin{equation*}
P_{n_7n_8}(\vartheta)=\langle n_7,n_8|\rho_{\vartheta}|n_7,n_8\rangle,
\end{equation*}
the related CFI is
\begin{equation}
F_C^{\mathrm{PNR}}(\vartheta)=\sum_{\substack{n_7,n_8\\P_{n_7n_8}>0}}
\frac{\left[\partial_{\vartheta}P_{n_7n_8}(\vartheta)\right]^2}{P_{n_7n_8}(\vartheta)}.
\label{eq:cfi_pnr_general}
\end{equation}
As the coincidence detection is a coarse graining of the complete photon-number measurement,
\begin{equation}
F_C^{\mathrm{coinc}}(\vartheta)\leq F_C^{\mathrm{PNR}}(\vartheta) \leq F_Q(\vartheta).
\label{eq:FI_hierarchy}
\end{equation}

The counting model above is in the case that the number of input trials is known. Using the zero-photon outcome requires heralded or otherwise clocked pair preparation, so that two-photon loss can be distinguished from the absence of an input pair. Throughout, we use the equal-resource normalization of Sec.~\ref{sec:reduced_detected_state}; $N_{\mathrm{rep}}$ counts independent trials and is distinct from the photon-number resource per trial.
In the case of $t(\delta)=0$, the phase $\arg[t(\delta)]$ is undefined, but the reduced state of the light is well defined. Thus the exact SLD QFI at this point is calculated directly from $\rho_{\vartheta}$ and $\partial_{\vartheta}\rho_{\vartheta}$. 
In zero transmission the rank of the reduced density matrix of the light changes. In addition  
the exact QFI value is distinguished from the limiting QFI obtained by approaching the zero through nearby regular points; see Sec.~\ref{sec:wgm_results} and Appendix~\ref{app:qfi_derivation}.

\subsection{Parameters of interest}

Using the polar decomposition of the transmission amplitude
of WGM that is introduced in Eq.~\eqref{eq:t_T_phi}, an unknown parameter $\vartheta$ may be encoded in
the intensity transmission $T$, the optical phase $\phi$, or both.
We focus on the angular-frequency detuning $\delta$ as a predominantly
dispersive parameter and the round-trip amplitude transmission
$\alpha$ as a parameter associated with intrinsic loss.

\subsubsection{WGM detuning}

The angular-frequency detuning is $\delta=\omega-\omega_0$, where $\omega$ is the probe  frequency and $\omega_0$ is the WGM resonance frequency. For fixed $\omega$, estimating a shift in $\omega_0$ is equivalent to estimating $-\delta$ and gives the same Fisher information.

Detuning has a dispersive physical origin, but changing $\delta$  modifies both $T(\delta)$ and $\phi(\delta)$, thus their full effects must be considered in the calculations. The dependence of resonance width to coupling is already contained in $t(\delta,\alpha,r)$, and we do not rescale the detuning by a linewidth that varies with $r$.
All analytical expressions use angular-frequency detuning. For plots in ordinary-frequency units, we define 
\begin{equation}
\delta_{\mathrm{MHz}} =\frac{\delta}{2\pi\times10^{6}\,\mathrm{s}^{-1}}.
\label{eq:delta_MHz}
\end{equation}
The corresponding Fisher informations satisfy
\begin{equation}
F_{Q,C}(\delta_{\mathrm{MHz}})=\left(2\pi\times10^{6}\,\mathrm{s}^{-1}\right)^2 F_{Q,C}(\delta).
\end{equation}

\subsubsection{Intrinsic loss and absorption}

The dissipative parameter investigated here is the WGM round-trip amplitude transmission $\alpha$. A reduction in $\alpha$ indicates increased intrinsic loss. If a sample introduces an additional intensity-attenuation coefficient $\Gamma_{\mathrm{abs}}$, while the background losses remain fixed, then
\begin{equation}
\alpha=\alpha_{\mathrm{bg}}\exp\!\left(-\frac{\Gamma_{\mathrm{abs}}L}{2}\right),
\label{eq:alpha_absorption}
\end{equation}
where $\alpha_{\mathrm{bg}}$ is the round-trip amplitude transmission in the absence of the additional absorption. The corresponding natural-log absorbance, or round-trip optical depth, is 
\begin{equation}
A \equiv -\ln\!\left(\frac{\alpha^2}{\alpha_{\mathrm{bg}}^2}\right)=\Gamma_{\mathrm{abs}}L.
\label{eq:absorbance_definition}
\end{equation}
If the background loss is not separated explicitly, one may set $\alpha_{\mathrm{bg}}=1$ and interpret $A=-\ln(\alpha^2)$ as the total round-trip optical depth. 

Fisher information transforms under a change of parameter according to \begin{equation}
F_{Q,C}^{(q)}=F_{Q,C}^{(\vartheta)}\left(\frac{d\vartheta}{dq}\right)^2.
\label{eq:FI_reparametrization}
\end{equation}
Applying this relation to Eq.~\eqref{eq:alpha_absorption} gives
\begin{equation}
\begin{aligned}
F_{Q,C}(\Gamma_{\mathrm{abs}})&=\frac{L^2\alpha^2}{4}F_{Q,C}(\alpha),\\
F_{Q,C}(A)&=\frac{\alpha^2}{4}F_{Q,C}(\alpha).
\end{aligned}
\label{eq:FI_absorption_conversion}
\end{equation}

The designation of $\alpha$ as a dissipative parameter refers to its physical origin. Because $\alpha$ appears inside the resonant transfer function in Eq.~\eqref{eq:WGM_transmission}, varying it can modify both $T$ and $\phi$, as defined in Eq.~\eqref{eq:t_T_phi}.

\section{General Origin of the Parameter-Selective Advantage}
\label{sec:parameter_selective_advantage}

\subsection{General QFI expressions}
\label{sec:general_qfi}

We now investigate how the relative performance of the two probes depends on the way in which a generic parameter $\vartheta$ is encoded in the complex WGM transmission amplitude in Eq.~\eqref{eq:t_T_phi} (having intensity transmission $T$ and phase $\phi$). The two different kinds of probes are compared in the equal-incident-photon number discussed in Sec.~\ref{sec:reduced_detected_state}.

Using the orthogonal decomposition of the reduced output state in Appendix~\ref{app:qfi_derivation}, and the invariance of the QFI under the parameter-independent second MZI beam splitter,  we obtain the photon-pair QFI 
\begin{equation}
F_Q^{\mathrm{pair}}(\vartheta) = \frac{\left(\partial_\vartheta T\right)^2}{T(1-T)}
+\frac{8T^2}{1+T^2}\left(\partial_\vartheta\phi\right)^2 .
\label{eq:qfi_pair_general}
\end{equation}
The equal-mean-photon-number coherent-state QFI is 
\begin{equation}
F_Q^{\mathrm{coh}}(\vartheta) = 4\left|\partial_\vartheta t\right|^2=
\frac{\left(\partial_\vartheta T\right)^2}{T}+ 4T\left(\partial_\vartheta\phi\right)^2 .
\label{eq:qfi_coherent_general}
\end{equation}
These expressions separate the information encoded 
in transmission from that associated with changes in phase. They
apply in the range  $0<T<1$; the rank-changing boundary
points are discussed separately in Sec.~\ref{sec:wgm_results} and Appendix~\ref{app:qfi_derivation}.

\subsection{Transmission- and phase-only limits}

For a parameter encoded only in the transmission magnitude,
we have $\partial_\vartheta\phi=0$, then the QFI expressions give
\begin{equation}
\frac{F_Q^{\mathrm{pair}}}{F_Q^{\mathrm{coh}}}
=\frac{1}{1-T}>1,
\qquad 0<T<1.
\label{eq:pure_loss_advantage}
\end{equation}
The advantage is due to the fixed photon number of the probe and
the parameter-dependent probabilities of the zero-, one-, and
two-photon loss sectors. The divergence happens at $T\rightarrow1$ and is a
one-sided local limit related to estimating transmissivity at the
boundary of its physical domain.

On the other hand, in phase-only encoding, $\partial_\vartheta T=0$ and  the
ratio is
\begin{equation}
\frac{F_Q^{\mathrm{pair}}}{F_Q^{\mathrm{coh}}}
=\frac{2T}{1+T^2}\leq1,
\label{eq:pure_phase_ratio}
\end{equation}
that equality is only in the lossless limit. Thus, the doubled two-photon
phase provides phase super-resolution relative to the single-photon phase response  but no advantage for the QFI .

As seen from Eq.~\eqref{eq:two_photon_output_state},
the surviving two-photon amplitude from the WGM arm is proportional to
$t^2=T e^{2i\phi}$. Therefore, this contribution vanishes as
$T\rightarrow0$, and a rapid phase variation near, or an undefined phase at,
a transmission zero cannot by itself produce a physical enhancement.

\subsection{Parameters encoded jointly in transmission and phase}

A general WGM parameter modifies both $T$ and $\phi$. Subtracting Eq.~\eqref{eq:qfi_coherent_general} from Eq.~\eqref{eq:qfi_pair_general} gives
\begin{equation}
F_Q^{\mathrm{pair}}-F_Q^{\mathrm{coh}}=\frac{(\partial_{\vartheta}T)^2}{1-T}
-\frac{4T(1-T)^2}{1+T^2}(\partial_{\vartheta}\phi)^2.
\label{eq:qfi_difference_general}
\end{equation}
The first term is non-negative and favours the photon-pair probe, whereas the second term is non-positive and favours the coherent-state probe. The photon pair provides an advantage precisely when
\begin{equation}
\frac{(\partial_{\vartheta}T)^2}{1-T}>\frac{4T(1-T)^2}{1+T^2}(\partial_{\vartheta}\phi)^2.
\label{eq:advantage_condition}
\end{equation}

This shows that the metrological efficiency depends on the parameter. A strong transmission response can produce a higher QFI for the photon-pair probe, whereas a parameter that is mainly encoded in phase does not benefit from the doubled two-photon phase. 
For the WGM parameters $\delta$ and $\alpha$, both contributions should be considered; thus, their complete derivatives must be computed using the physical transfer function $t(\delta,\alpha,r)$.

\section{WGM Parameter-Estimation Results}
\label{sec:wgm_results}

We now apply the general results of Sec.~\ref{sec:parameter_selective_advantage} to the WGM transmission amplitude of Eq.~\eqref{eq:WGM_transmission}, using the round-trip phase $\theta_{\mathrm{WGM}}(\delta)=\tau_{\mathrm{rt}}\delta$ defined in Eq.~\eqref{eq:theta_wgm}, where $\tau_{\mathrm{rt}}=nL/c$ and the negligible material dispersion assumption. For all numerical results, we use a resonator radius $R=50~\mu\mathrm{m}$ and an effective refractive index $n=1.45$, corresponding to $\tau_{\mathrm{rt}}=n(2\pi R)/c$. Unless stated otherwise, the round-trip amplitude transmission is taken to be $\alpha=0.9997$. For a resonance wavelength near 810~nm, this value of $\alpha$ corresponds at critical coupling to an intensity linewidth (FWHM) of approximately 275~fm, or 126~MHz (equivalently, an HWHM of approximately 63~MHz; the relation between WGM linewidth and parameters $\alpha, r$ is given in Appendix~\ref{app:wgm_linewidth}). These parameters were chosen to approximately match experimental values for WGM single molecule sensing \cite{Vincent_Subramanian2020, Subramanian_Vincent2020}, and a wavelength at which photon pairs are commonly produced. Writing $D(\delta,\alpha,r)=1-r\alpha e^{i\theta_{\rm WGM}(\delta)}$, the required derivatives are 
\begin{equation}
\partial_{\delta}t=-\frac{i\tau_{\mathrm{rt}}\alpha(1-r^2)e^{i\theta_{\mathrm{WGM}}}}{D^2},\quad
\partial_{\alpha}t=-\frac{(1-r^2)e^{i\theta_{\mathrm{WGM}}}}{D^2}.
\label{eq:wgm_t_derivatives}
\end{equation}
The corresponding derivatives of the intensity transmission and phase are obtained from
\begin{equation}
\partial_{\vartheta}T=2\operatorname{Re}\left[t^*\partial_{\vartheta}t\right],\quad
\partial_{\vartheta}\phi=\frac{\operatorname{Im}\left[t^*\partial_{\vartheta}t\right]}{T},
\label{eq:T_phi_derivatives}
\end{equation}
for $\vartheta\in\{\delta,\alpha\}$ and $T>0$. At zero transmission, the QFI is evaluated directly from the reduced state rather than from the second expression in Eq.~\eqref{eq:T_phi_derivatives}.

\subsection{Detuning estimation}

At zero detuning, the WGM transmission amplitude is real:
\begin{equation}
t_0 \equiv t (0,\alpha,r)=\frac{r-\alpha}{1-r\alpha},\quad
T_0=\left(\frac{r-\alpha}{1-r\alpha}\right)^2.
\label{eq:resonant_transmission}
\end{equation}
For $r\neq\alpha$, the first-order transmission response vanishes at resonance, $\left.\partial_\delta T\right|_{\delta=0}=0$, whereas $\partial_{\delta}\phi$ is generally nonzero. Thus the detuning is encoded to first order through the phase at the center of the resonance. The coherent-state QFI at this point is
\begin{equation}
F_Q^{\mathrm{coh}}(\delta;0)=\frac{4\tau_{\mathrm{rt}}^2\alpha^2(1-r^2)^2}{(1-r\alpha)^4}.
\label{eq:coh_detuning_resonance}
\end{equation}
Using Eq.~\eqref{eq:pure_phase_ratio}, we have
\begin{equation}
\frac{F_Q^{\mathrm{pair}}(\delta;0)}{F_Q^{\mathrm{coh}}(\delta;0)}=\frac{2T_0}{1+T_0^2}\leq1.
\label{eq:detuning_resonance_ratio}
\end{equation}
Thus, at the phase-dominated centre of a regular resonance ($r\neq\alpha$), the doubled two-photon phase does not provide an advantage over the coherent-state probe.

In the regions far from the zero detuning, both $\partial_{\delta}T$ and $\partial_{\delta}\phi$ contribute. Local regions with $F_Q^{\mathrm{pair}}>F_Q^{\mathrm{coh}}$ may then occur when the transmission contribution satisfies Eq.~\eqref{eq:advantage_condition}.

This enhancement is due to the detuning-dependent loss statistics, i.e., the detuning dependence of the zero-, one-, and two-photon loss probabilities,  instead of phase super-resolution.
Since a large ratio may happen where both QFIs are small, the practical advantage must be checked together with the absolute QFI.

\begin{figure*}[t]
\centering
\subfloat[]{%
\label{fig:detuning_qfi_r0999}%
\includegraphics[width=0.32\textwidth]{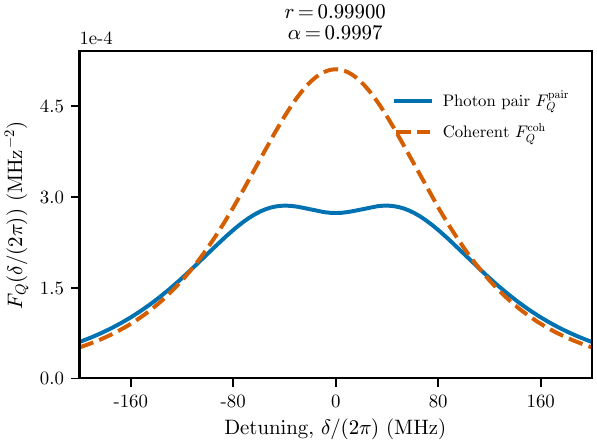}}%
\hfill
\subfloat[]{%
\label{fig:detuning_qfi_r099968}%
\includegraphics[width=0.32\textwidth]{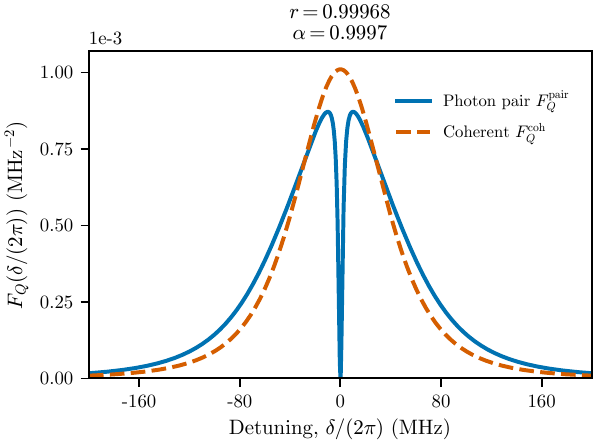}}%
\hfill
\subfloat[]{%
\label{fig:detuning_qfi_r09998}%
\includegraphics[width=0.32\textwidth]{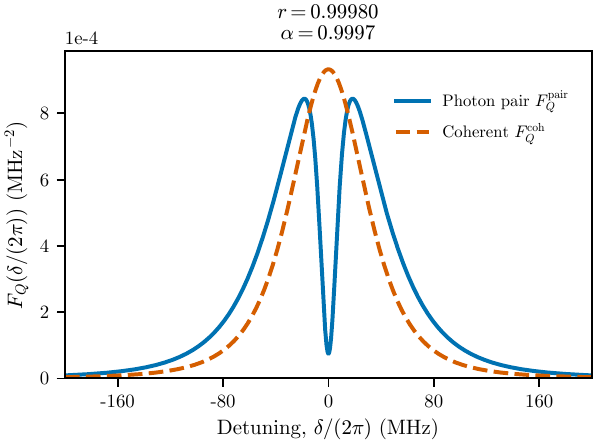}}%
\caption{Quantum Fisher information for estimation of the WGM detuning, comparing the photon-pair probe (solid blue curves) with the coherent-state probe (dashed orange curves) at the same mean incident photon number, $\overline{N}_{\mathrm{in}}=2$. The round-trip amplitude transmission is fixed at $\alpha=0.9997$, the coupling amplitudes are  (a) $r=0.99900$, (b) $r=0.99968$, and (c) $r=0.99980$. The detuning is expressed as $\delta/(2\pi)$ in MHz, and the QFIs are evaluated for this frequency variable and have units of $\mathrm{MHz}^{-2}$. At the resonance centre, detuning is encoded to first order through the phase response and the photon-pair probe does not outperform the coherent-state case.}
\label{fig:detuning_qfi}
\end{figure*}
Figure~\ref{fig:detuning_qfi} shows the QFI values using photon-pair and coherent-state probes at the same physical detuning and the same mean photon number, for three different coupling values used in Fig.~\ref{fig:wgm_transmission_coinc_a0p9997_varying_r}. At zero detuning, the numerical results confirm the phase-dominated relation mentioned above. For $r=0.99900$, $0.99968$, and $0.99980$, the ratios $F_Q^{\mathrm{pair}}/F_Q^{\mathrm{coh}}$ at resonance are approximately $0.535$, $0.0021$, and $0.080$, respectively. The strong suppression near critical coupling is because of very small $T_0$ values.

At exact critical coupling, $r=\alpha$ and $\delta=0$, the photon-pair reduced state changes only to second order in a small detuning displacement, and we have $\partial_{\delta}\rho_{\rm pair}=0$ for zero transmission values. Consequently, the standard SLD QFI of the photon-pair state at the exact transmission zero is zero.  In the case of non-zero detuning, the previously unoccupied one-photon-loss sector is populated with a probability of order $\delta^{2}$, thus the QFI can have a finite limit as $\delta\rightarrow0$ through regular points. The exact QFI at $\delta=0$ differs from the limiting value obtained as $\delta\rightarrow0$ through nonzero detuning values. 
In addition, the limiting QFI may depend on the direction taken to the critical point, which indicates the statistical model is nonregular there.

\subsection{Intrinsic-loss estimation}

Here $F_Q^{(s)}(\alpha;0)$ shows the QFI for estimating $\alpha$ when detuning is zero ($\delta=0$) and coupling $r$ is fixed, with $s\in\{\mathrm{pair},\mathrm{coh}\}$.

We now consider the amplitude transmission $\alpha$  unknown, with fixed $r$ and the frequency of the probe. In the case of zero detuning, both $t_0$ and $\left.\partial_\alpha t\right|_{\delta=0}$ are real valued; thus $\left.\partial_\alpha\phi\right|_{\delta=0}=0$ in $r\neq\alpha$, and $\alpha$ is locally encoded in the transmission magnitude. The QFI of the coherent-state is 
\begin{equation}
F_Q^{\mathrm{coh}}(\alpha;0)=\frac{4(1-r^2)^2}{(1-r\alpha)^4},
\label{eq:coh_alpha_resonance}
\end{equation}
whereas the photon-pair QFI is
\begin{equation}
F_Q^{\mathrm{pair}}(\alpha;0)=\frac{4(1-r^2)}{(1-\alpha^2)(1-r\alpha)^2}.
\label{eq:pair_alpha_resonance}
\end{equation}
Their ratio is
\begin{equation}
\mathcal{E}_{\alpha}(r,0)\equiv \frac{F_Q^{\mathrm{pair}}(\alpha;0)}{F_Q^{\mathrm{coh}}(\alpha;0)}=\frac{1}{1-T_0}= \frac{(1-r\alpha)^2}{(1-r^2)(1-\alpha^2)}>1 .
\label{eq:alpha_enhancement_resonance}
\end{equation}

For every regular noncritical point in the range of $0<T_0<1$, the photon-pair shows a genuine enhancement over the coherent-state benchmark for estimating intrinsic WGM attenuation. 

\begin{figure*}[t]
\centering
\subfloat[]{%
\label{fig:alpha_qfi_r0999}%
\includegraphics[width=0.32\textwidth]{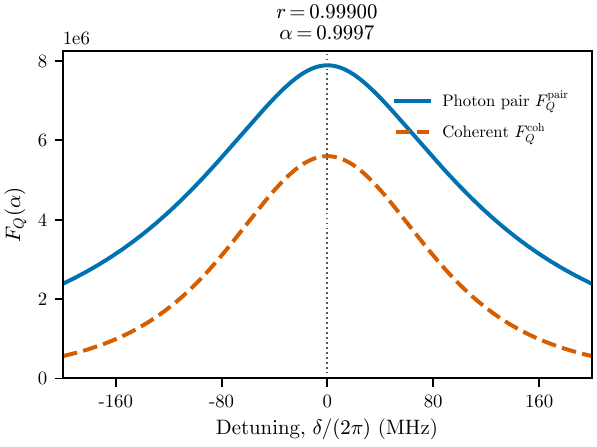}}%
\hfill
\subfloat[]{%
\label{fig:alpha_qfi_r099968}%
\includegraphics[width=0.32\textwidth]{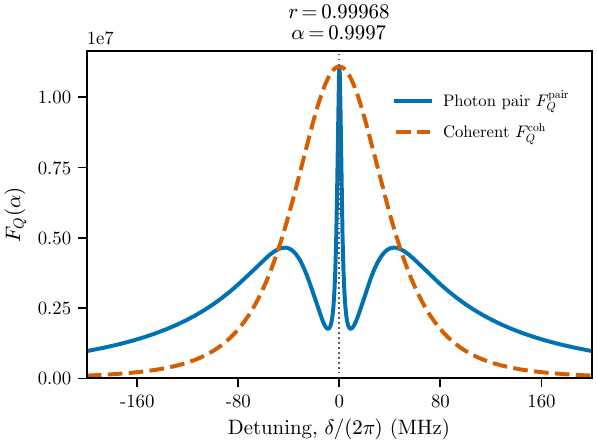}}%
\hfill
\subfloat[]{%
\label{fig:alpha_qfi_r09998}%
\includegraphics[width=0.32\textwidth]{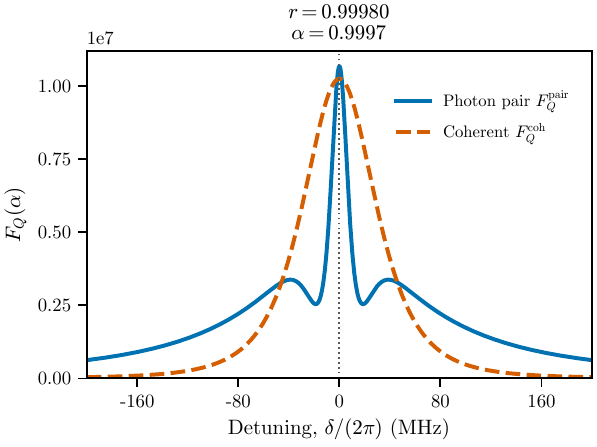}}%
\caption{QFI for estimation of the WGM round-trip amplitude transmission $\alpha$, comparison of the photon-pair (solid blue) and equal-mean-photon-number coherent-state (dashed orange) probes for (a) $r=0.99900$, (b) $r=0.99968$, and (c) $r=0.99980$, with $\alpha=0.9997$. The vertical dotted line indicates zero detuning, and the enhancement of the photon-pair  away from exact value of the critical coupling is $F_Q^{\mathrm{pair}}/F_Q^{\mathrm{coh}}=1/(1-T_0)$. Off resonance, both transmission and phase variations have contributions in the QFI.}
\label{fig:alpha_qfi}
\end{figure*}

Figure~\ref{fig:alpha_qfi} extends the resonant analytical comparison to nonzero detuning. At $\delta=0$, the photon-pair probe outperforms the coherent-state probe at each of the three regular operating points, in agreement with
$\mathcal{E}_\alpha(r,0)=1/(1-T_0)$. The corresponding enhancement factors are approximately $1.409$, $1.001$, and $1.042$ for $r=0.99900$, $0.99968$, and $0.99980$, respectively. Away from resonance, changes in $\alpha$ affect both $T$ and $\phi$, and the phase contribution can cause the coherent-state QFI to exceed the photon-pair QFI over parts of the spectrum.

\subsection{Enhancement versus absolute sensitivity}

The enhancement ratio alone does not clarify what is the best sensing configuration. At zero detuning and fixed $\alpha$, the coherent-state QFI attains its maximum at $r=\alpha$. By contrast, Eq.~\eqref{eq:pair_alpha_resonance} for the QFI of the photon-pair is only valid for regular points with $r\neq\alpha$. Over the regular operating points, its supremum is equal to the coherent-state maximum:
\begin{equation}
\begin{aligned}
\sup_{r\neq\alpha}F_Q^{\rm pair}(\alpha;0) &= \lim_{\substack{r\rightarrow\alpha\\ r\neq\alpha}}F_Q^{\rm pair}(\alpha;0) \\
&=\max_{r}F_Q^{\rm coh}(\alpha;0)= \frac{4}{(1-\alpha^{2})^{2}}.
\end{aligned}
\label{eq:qfi_critical_supremum}
\end{equation}
At the same time,
\begin{equation}
\lim_{\substack{r\rightarrow\alpha\\r\neq\alpha}}
\mathcal{E}_{\alpha}(r,0)=1.
\end{equation}
In order to explain the nonregular behaviour explicitly we set $r=\alpha+\epsilon$ at fixed $\alpha$ and zero detuning. Then $t_0=\epsilon/(1-\alpha^2)+O(\epsilon^2)$ and Eq.~\eqref{eq:alpha_enhancement_resonance} gives $\mathcal{E}_{\alpha}(\alpha+\epsilon,0)=1+\epsilon^2/(1-\alpha^2)^2+O(\epsilon^3)$, showing why the enhancement tends to unity. At the exact critical point, varying the estimated parameter as $\alpha\to\alpha+\eta$ while $r$ is constant gives $t_0=-\eta/(1-\alpha^2)+O(\eta^2)$. The reduced photon-pair state varies only with $|t_0|^2$, thus $\partial_\alpha\rho_{\rm pair}=0$ and its exact SLD QFI vanishes, although we have the finite regular-point limit in Eq.~\eqref{eq:qfi_critical_supremum}.

The supremum in  Eq.~\eqref{eq:qfi_critical_supremum} is calculated over regular, noncritical operating points, and does not correspond to photon-pair SLD QFI at the  $r=\alpha$ point. At  $r=\alpha$, we have $\partial_{\alpha}\rho_{\rm pair}=0$; thus, the value of photon-pair SLD QFI is zero. On the other hand the coherent-state QFI is finite since $\partial_{\alpha}t\neq0$ and the output of the coherent state changes to first order at $\alpha-r$.

Far from critical coupling the ratio $\mathcal{E}_{\alpha}$ is increased but the absolute QFI is reduced. In the weak-coupling regime where $r\rightarrow1$, the enhancement ratio can be large and the interaction with the resonator, as well as the absolute information, vanishes.

The photon-pair advantage occurs for a fixed coupling condition, and is not at critical coupling. It is also robust for small fluctuations in $r$ or $\alpha$. 
If we can select $r$  freely for each probe and parameter value, the best coherent-state QFI is the same as the limiting maximum photon-pair QFI. The enhancement map must be considered  with the absolute photon-pair QFI to identify the advantageous regions.

\begin{figure*}[t]
\centering
\subfloat[]{%
\label{fig:alpha_enhancement_map_a}%
\includegraphics[width=0.48\textwidth]{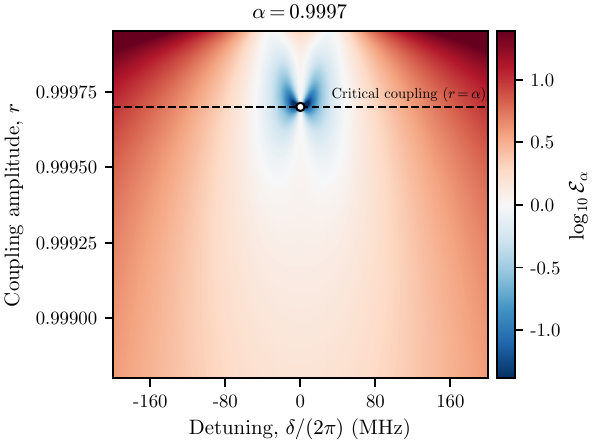}}%
\hfill
\subfloat[]{%
\label{fig:alpha_enhancement_map_b}%
\includegraphics[width=0.48\textwidth]{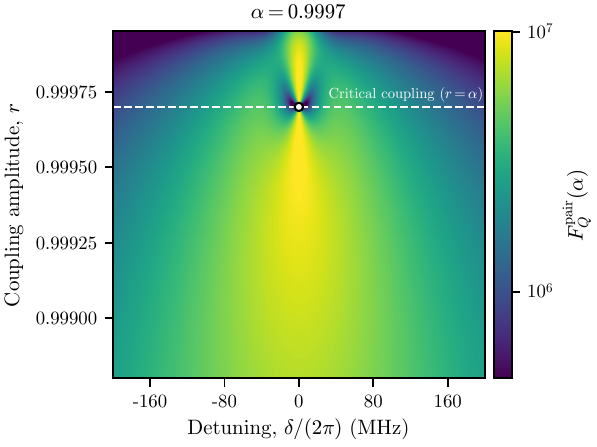}}%
\caption{Photon-pair advantage in QFI for estimation of the WGM round-trip amplitude transmission $\alpha$, as a function of coupling amplitude $r$ and detuning $\delta/(2\pi)$. The photon-pair and coherent-state probes have the same mean photon number, $\overline{N}_{\mathrm{in}}=2$. The reference value is $\alpha=0.9997$. Panel (a) is the logarithm of the QFI enhancement factor, $\log_{10}\mathcal{E}_{\alpha}(r,\delta)$, where $\mathcal{E}_{\alpha}(r,\delta)=F_Q^{\mathrm{pair}}(\alpha)/F_Q^{\mathrm{coh}}(\alpha)$. Red areas correspond to a photon-pair advantage, $\mathcal{E}_{\alpha}(r,\delta)>1$, while blue areas correspond to $\mathcal{E}_{\alpha}(r,\delta)<1$. Panel (b) is the absolute photon-pair QFI, $F_Q^{\mathrm{pair}}(\alpha)$, plotted on a logarithmic colour scale. The dashed horizontal line indicates critical coupling, $r=\alpha$. The open white marker at $r=\alpha$ and $\delta=0$ denotes the exact critical-coupling transmission zero. This point is masked because it is a nonregular statistical point and therefore the white marker should not be interpreted as a value of $\log_{10}\mathcal{E}_{\alpha}(r,\delta)$. The surrounding colour-map values correspond to regular points arbitrarily close to the transmission zero. Along regular points approaching the critical point at zero detuning, $\mathcal{E}_{\alpha}(r,0)\rightarrow1$. The trade-off between relative enhancement and absolute sensitivity shows that large enhancement ratios do not necessarily correspond to large absolute QFI.}
\label{fig:alpha_enhancement_map}
\end{figure*}

Figure~\ref{fig:alpha_enhancement_map} shows the trade-off between relative enhancement and absolute QFI. Panel (a) identifies regions with $\mathcal{E}_{\alpha}(r,\delta)>1$, and (b) shows the absolute photon-pair QFI. Large enhancement ratios can occur where the absolute information is small, while the largest absolute QFI is concentrated near resonance and critical coupling. Along regular points approaching critical coupling at zero detuning, $\mathcal{E}_{\alpha}\rightarrow 1$. A useful operating point must therefore combine $\mathcal{E}_{\alpha}>1$ with sufficiently large $F_Q^{\mathrm{pair}}(\alpha)$. The exact point $r=\alpha,\delta=0$ is nonregular and is excluded from the colour map.

\subsection{Measurement-accessible information}
\label{sec:measurement_information}
At regular noncritical resonance points, the advantage in Eq.~\eqref{eq:alpha_enhancement_resonance} is experimentally accessible through photon-number-resolved detection, provided that the photon-pair preparation trials are heralded or otherwise clocked, so that the vacuum outcome associated with two-photon loss can be distinguished from the absence of an input pair. At zero detuning, $t_0$ is real and the output probabilities depend only on $T_0$. Direct substitution of these probabilities into Eq.~\eqref{eq:cfi_pnr_general} gives
\begin{equation}
F_C^{\mathrm{PNR}}(\alpha;0)=F_Q^{\mathrm{pair}}(\alpha;0).
\label{eq:pnr_saturates_alpha_qfi}
\end{equation}
Thus, at regular noncritical resonance points with $0<T_0<1$, ideal PNR detection saturates the photon-pair QFI for intrinsic-loss estimation. Direct photon counting at both coherent-state output ports similarly saturates the coherent-state QFI at this operating point.

Using the coincidence probability in Eq.~\eqref{eq:P11} in the binary CFI defined in Eq.~\eqref{eq:cfi_coinc_general}, the information retained by coincidence-only detection at resonance satisfies
\begin{equation}
\frac{F_C^{\mathrm{coinc}}(\alpha;0)}{F_Q^{\mathrm{pair}}(\alpha;0)}=\frac{4T_0}{T_0+3}\leq1.
\label{eq:coincidence_fraction_alpha}
\end{equation}
Coincidence detection becomes particularly inefficient close to a transmission zero, where the information is predominantly carried by the other photon-number outcomes. PNR detection is therefore the natural measurement for exploiting the loss-estimation advantage.

\subsection{Unresolved detuning fluctuations}
\label{sec:unresolved_detuning}
To model a common detuning offset that is unknown within each measurement trial and varies randomly between trials, we write the actual detuning as $\delta=\delta_{0}+\Delta$. We assume that $\Delta$ is drawn from a parameter-independent, zero-mean Gaussian distribution. This model can represent unresolved shot-to-shot resonance-frequency jitter or an incoherent common-mode variation of the probe frequency. The physical state must be averaged over this distribution before calculating the QFI:
\begin{equation}
\overline{\rho}_{\vartheta}(\delta_{0}) =\int_{-\infty}^{\infty}d\Delta\,K_{\sigma}(\Delta) \rho_{\vartheta}(\delta_{0}+\Delta),
\label{eq:averaged_state_general}
\end{equation}
where
\begin{equation}
K_{\sigma}(\Delta)=\frac{1}{\sqrt{2\pi}\sigma} \exp\left(-\frac{\Delta^{2}}{2\sigma^{2}}\right).
\end{equation}
The effective QFI is then $F_{Q,\mathrm{eff}}(\vartheta;\delta_0)=F_Q\!\left[\overline{\rho}_{\vartheta}(\delta_0)\right]$.
Similarly, measurement probabilities must first be averaged,
\begin{equation}
\overline{P}_{n_7n_8}(\delta_0)=\int_{-\infty}^{\infty}d\Delta\,K_{\sigma}(\Delta)P_{n_7n_8}(\delta_0+\Delta),
\end{equation}
before evaluating the corresponding CFI. Averaging an ideal Fisher information is not the same as the accessible Fisher information.
All effective quantities discussed below refer to the classical detuning mixture defined in Eq.~\eqref{eq:averaged_state_general}. This model is not a coherent finite-bandwidth photon-pair wavepacket. For obtaining the latter, propagation of the complete joint spectral amplitude is required, followed by tracing over the unresolved frequency degrees of freedom. This is not considered here.

\begin{figure*}[t]
\centering
\subfloat[]{%
\label{fig:alpha_finite_width_r0999}%
\includegraphics[width=0.32\textwidth]{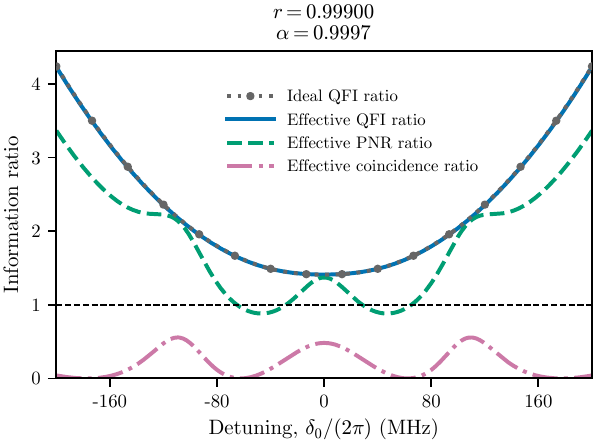}}%
\hfill
\subfloat[]{%
\label{fig:alpha_finite_width_r099968}%
\includegraphics[width=0.32\textwidth]{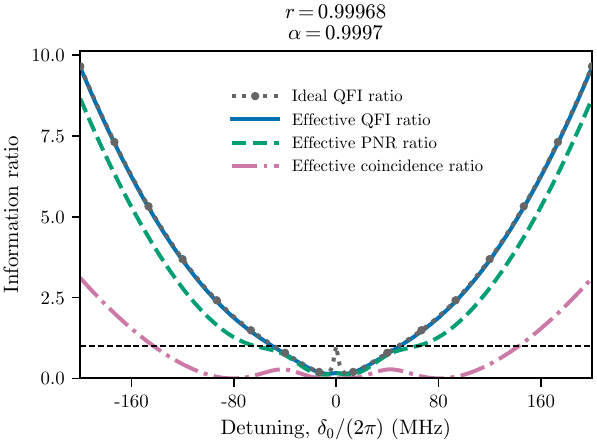}}%
\hfill
\subfloat[]{%
\label{fig:alpha_finite_width_r09998}%
\includegraphics[width=0.32\textwidth]{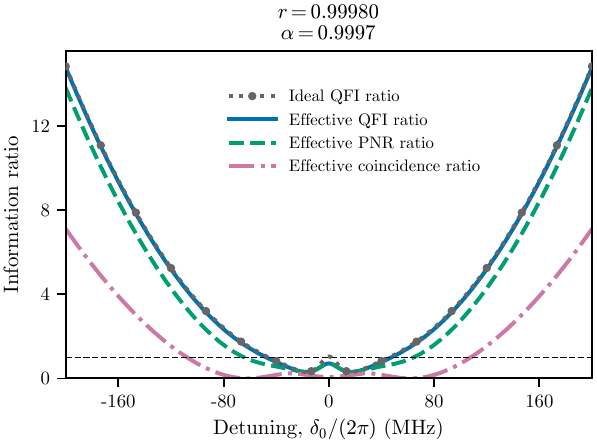}}%
\caption{Effect of unresolved Gaussian detuning offsets on estimation of the WGM round-trip amplitude transmission $\alpha$. Independent source and resonance-jitter contributions with $\sigma_{\mathrm{src}}/(2\pi)=6.3~\mathrm{MHz}$ and $\sigma_{\mathrm{jit}}/(2\pi)=0.46~\mathrm{MHz}$ give us $\sigma_{\mathrm{eff}}/(2\pi)\simeq6.32~\mathrm{MHz}$. The grey dotted, solid blue, dashed green, and dash-dotted magenta curves show the ideal QFI, effective QFI, effective PNR-CFI, and effective coincidence-CFI ratios, respectively. They are normalized by the coherent-state QFI in the related ideal or averaged model. States and probabilities are averaged before evaluating the information. The horizontal dashed line marks equality with the coherent-state benchmark. At zero detuning, the photon-pair advantage survives for $r=0.99900$ and is nearly saturated by PNR detection, but it is absent at the two operating points closer to critical coupling.}
\label{fig:alpha_finite_width}
\end{figure*}

Figure~\ref{fig:alpha_finite_width} compares the ideal and effective information ratios after averaging the physical states and measurement probabilities over the unresolved detuning distribution. We take the total angular-frequency offset to be $\Delta=\Delta_{\rm src}+\Delta_{\rm jit}$, where the two contributions are independent, zero-mean Gaussian random variables. Their angular-frequency standard deviations satisfy $\sigma_{\rm eff}=\sqrt{\sigma_{\rm src}^{2}+\sigma_{\rm jit}^{2}}$. For the present model, we take $\sigma_{\mathrm{src}}/(2\pi)=6.3~\mathrm{MHz}$ and $\sigma_{\mathrm{jit}}/(2\pi)=0.46~\mathrm{MHz}$ as representative rms widths, giving $\sigma_{\mathrm{eff}}/(2\pi)\simeq6.32~\mathrm{MHz}$. These representative widths are motivated by experimentally demonstrated WGM spectral resolution at the $\sim1$~fm scale \cite{Toropov_Houghton2024, Arunkumar_Zossimova2025} and narrowband photon-source experiments \cite{Wolfgramm_Astiz2011, Fekete_Rielander2013}. In the present detuning-mixture model, $\sigma_{\mathrm{src}}$ represents an effective rms detuning spread of approximately 10\% of the WGM half-width at half maximum (HWHM) at critical coupling. This averaging suppresses the narrow structures near the critical point. The wavy structure of the coincidence curve for $r=0.99900$ comes from the interference term in $P_{11}\propto |1+t^2|^2$. Closer to critical coupling, these spectral features are narrower and are more strongly smoothed by the detuning averaging.

At zero nominal detuning, the QFI ratios are nearly $1.410$, $0.167$, and $0.713$ when we use $r=0.99900$, $0.99968$, and $0.99980$, respectively. The related PNR information ratios are $1.372$, $0.163$, and $0.712$. Then, for this unresolved width, a practically accessible photon-pair advantage persists at $r=0.99900$; in contrast, it does not exist for the near-critical and undercoupled cases. Since coincidence-only detection discards the one- and two-photon loss outcomes, it remains below the coherent-state QFI case at resonance in all three cases.

\section{Implications for Absorption and Pump--Probe Sensing}
\label{sec:absorption_applications}

\subsection{Absorption and absorbance sensing}

The result obtained for the intrinsic-loss estimation earlier can be used for estimation of experimentally relevant absorption quantities, such as the absorption coefficient and absorbance. We consider a sample parameter $x$ that changes the WGM response mainly via an additional attenuation coefficient $\Gamma_{\mathrm{abs}}(x)$. Using the relation between $\alpha$ and $\Gamma_{\mathrm{abs}}$ in Eq.~\eqref{eq:alpha_absorption}, and the reparametrization rule in Eqs.~\eqref{eq:FI_reparametrization} and \eqref{eq:FI_absorption_conversion}, we have
\begin{equation}
F_{Q,C}^{(s)}(x) =\frac{L^2\alpha^2}{4}[\partial_x\Gamma_{\mathrm{abs}}(x)]^2F_{Q,C}^{(s)}(\alpha) \quad
s\in\{\mathrm{pair},\mathrm{coh}\}
\end{equation}
Provided $\partial_x\Gamma_{\mathrm{abs}}\neq0$, the ratio of the photon-pair-to-coherent state QFI is identical to that for $\alpha$; the same conclusion holds for the round-trip absorbance $A=\Gamma_{\mathrm{abs}}L$.

The reparametrization factor determines the absolute sensitivity and should be considered when we compare different samples. A large enhancement ratio does not necessarily imply large absolute QFI if $\partial_x\Gamma_{\mathrm{abs}}$ is small. This result assumes that the parameter mainly changes the absorption. If it also affects the resonance frequency, both the absorptive and dispersive contributions must be considered.

\subsection{Pump--probe readout of state-dependent absorption}

The same framework can be used for a pump--probe measurement in which a pump pulse excites the sample and the photon-pair probe measures the resulting absorption. In a simple two-state model, $p_e(\tau)$ is the excited-state population at pump--probe delay $\tau$. The effective absorption coefficient is
\begin{equation}
\Gamma_{\mathrm{abs}}(p_e) =\left(1-p_e\right)\Gamma_{\mathrm{abs}}^{(g)}+p_e\Gamma_{\mathrm{abs}}^{(e)} = \Gamma_{\mathrm{abs}}^{(g)}+p_e\Delta\Gamma_{\mathrm{abs}},
\end{equation}
where $\Delta\Gamma_{\mathrm{abs}} = \Gamma_{\mathrm{abs}}^{(e)} - \Gamma_{\mathrm{abs}}^{(g)}$. The QFI for estimating the population for a fixed delay is
\begin{equation}
F_Q^{(s)}(p_e) = \frac{L^{2}\alpha^{2}}{4} \left(\Delta\Gamma_{\mathrm{abs}}\right)^{2}F_Q^{(s)}(\alpha).
\end{equation}
Thus, when the pump-induced signal is mainly absorptive, we have
\begin{equation}
\frac{F_Q^{\mathrm{pair}}(p_e)}{F_Q^{\mathrm{coh}}(p_e)} =
\frac{F_Q^{\mathrm{pair}}(\alpha)}{F_Q^{\mathrm{coh}}(\alpha)}.
\end{equation}

The photon-pair advantage for intrinsic-loss estimation can also apply to transient population or differential absorbance.

In a fixed-coupling WGM resonator, such as a lithographically defined on-chip ring, a weakly absorbing molecule or nanoparticle can modify the intrinsic loss. The photon-pair advantage identified in Eq.~\eqref{eq:alpha_enhancement_resonance} may therefore be relevant to photon-limited pump--probe measurements, where reducing optical exposure and potential photodamage is important.

A pump-induced molecular change may also modify the refractive index and consequently the resonance detuning. In this more general case, $\alpha=\alpha(p_e)$ via the absorption that depends on state, while the dispersive response may similarly give $\delta=\delta(p_e)$. At fixed coupling amplitude $r$, the total derivative of the WGM transmission amplitude is
\begin{equation}
\frac{d t}{d p_e}=\left(\partial_\alpha t\right)\frac{d\alpha}{d p_e} + \left(\partial_\delta t\right)\frac{d\delta}{d p_e}.
\end{equation}
Thus the full transmission and phase contributions discussed in Sec.~\ref{sec:parameter_selective_advantage} must be kept. Advantage of the photon-pair QFI is possible only when the dissipative contribution is strong enough to satisfy Eq.~\eqref{eq:advantage_condition}. The PNR requirement in Sec.~\ref{sec:measurement_information}  applies to the pump--probe case as well. A model for a specific sample would need its absorption cross sections, state populations, and relaxation dynamics, which are not included here.

\section{Conclusion}
\label{sec:conclusion}

We have investigated the metrological performance of an indistinguishable photon-pair probe in a lossy WGM-coupled Mach--Zehnder interferometer. The physical output state was obtained by retaining the zero-, one-, and two-photon loss sectors generated by the WGM interaction and tracing only over the unobserved environmental mode. This treatment allows a direct comparison with a coherent-state probe under the same mean incident photon-number constraint.

The analytical QFI expressions reveal that the relative performance is determined by how the unknown parameter is encoded in the complex transmission amplitude. For pure phase estimation, the photon-pair-to-coherent QFI ratio is $2T/(1+T^2)\leq1$. The doubled phase of the surviving two-photon component therefore produces phase super-resolution but not a metrological advantage. For pure transmission estimation, the corresponding ratio is $1/(1-T)>1$, and the enhancement originates from the parameter dependence of the different photon-loss probabilities.

In estimating WGM detuning, in the regular resonance the response is phase dominated thus the photon-pair probe does not have higher QFI over the coherent-state. 
Away from the WGM resonance, a photon-pair may have advantage because of the transmission contribution, not the doubled phase.
On the other hand estimating the round-trip amplitude transmission $\alpha$ at resonance is locally dissipative and shows an advantage for photon-pair at fixed noncritical coupling. The enhancement ratio and absolute QFI are largest in different regions: near critical coupling the absolute QFI increases, while the enhancement ratio approaches one. If the coupling is optimized for each probe, the best coherent-state QFI equals the largest photon-pair QFI approached at regular noncritical points. This value is approached as $r\rightarrow\alpha$, but it is not the exact photon-pair SLD QFI at the transmission zero.

The loss-estimation advantage can be accessed with PNR detection, which keeps information from all photon-loss outcomes. Coincidence-only detection loses some of this information, especially near zero transmission. Detuning fluctuations smooth the narrow near-critical features. For the width considered here, the PNR advantage remains in the overcoupled regime but not near critical coupling.

Because Fisher information transforms directly under a change of parameter, the enhancement obtained for $\alpha$ also applies to absorption, absorbance, and state-dependent attenuation, provided that the signal is predominantly dissipative. This provides a route towards photon-pair-enhanced absorption and pump--probe sensing with WGM resonators. A complete treatment of a coherent finite-bandwidth photon-pair wavepacket, detector inefficiency, dark counts, imperfect photon indistinguishability, and source-heralding efficiency remains important for assessing the achievable advantage in a specific experimental implementation.

\begin{acknowledgments}
The authors would like to thank Professor Jos\'e A. Roversi and Dr Charles Downing for valuable discussions about the results and manuscript.

C.J. would like to thank Dr Lucia Caspani and Dr Isaac Luxmoore for their feedback on an early version of this manuscript, and Dr Jes\'us Rubio for advice about quantum metrology.

C.J. and F.V. acknowledge support from the Engineering and Physical Sciences Research Council (EPSRC), under grants EP/X018822/1 and EP/R031428/1.
F.V. acknowledges support from the University of Exeter
through QUBIC (CE230100021).
E.D. acknowledges funding by the Higher Education and Science Committee of Armenia under Grant No. 22IRF-06. 

A.V.-B. would like to thank the Air Force Office of Scientific Research (AFOSR), USA, under award N${\textsuperscript{\underline{o}}}$ FA9550-24-1-0009, and Conselho Nacional de Desenvolvimento Científico e Tecnol\'ogico, (CNPq), Brazil, via the Instituto Nacional de Ci\^encia e Tecnologia - Informa\c c\~ao Qu\^antica (INCT-IQ), grant N${\textsuperscript{\underline{o}}}$ 465469/2014-0.
\end{acknowledgments}

\bibliographystyle{apsrev4-2}
\bibliography{REFS}

\appendix

\section{Photon Pair Output State Derivation}
\label{app:photon_pair_derivation}

The relation between the input modes $(\hat a_1,\hat a_2)$ and the intermediate output modes $(\hat a_5,\hat a_6)$ shown in Fig.~\ref{fig:WGM_MZI_schematic} can be written as
\begin{equation}
    \begin{pmatrix}
        \hat{a}_5\\
        \hat{a}_6
    \end{pmatrix} = \dfrac{1}{\sqrt{2}} \begin{pmatrix}
        i t(\delta) & t(\delta)\\
        1 & i
    \end{pmatrix} \begin{pmatrix}
        \hat{a}_1\\
        \hat{a}_2
    \end{pmatrix} + \begin{pmatrix}
        \hat{F}(\omega)\\
        0
    \end{pmatrix}
    \label{eq:input-output_matrix}
\end{equation}

After the first beam splitter, the input photon pair forms a two-photon N00N state. To describe propagation through the lossy WGM, we define $\ell(\delta)=\sqrt{1-|t(\delta)|^{2}}$. For completeness, the OVPA noise operator appearing above can be expressed in terms of the distributed environmental modes $\hat{s}(z,\omega)$ as~\cite{Alsing_Hach2017}
\begin{equation}
\begin{aligned}
&\hat F(\omega)=-i|\kappa|^{2}\sqrt{\Gamma(\omega)}\sum_{m=0}^{\infty}(r^{*})^{m}\\
&\quad \times\int_{0}^{(m+1)L}dz\,\exp\!\left\{i\xi(\omega)\big[(m+1)L-z\big] \right\}\hat s(z,\omega),
\end{aligned}
\label{eq:noise_operator}
\end{equation}
where $\xi(\omega)=n(\omega)\omega/c+i\Gamma(\omega)/2$, and $\Gamma(\omega)$ is the intensity attenuation coefficient per unit length. This operator satisfies $[\hat F(\omega),\hat F^\dagger(\omega')] =\ell^2(\delta)\delta_D(\omega-\omega')$. It can therefore be written as $\hat F(\omega)=\ell(\delta)\hat b_{\mathrm{in}}(\omega)$, where $\hat b_{\mathrm{in}}(\omega)$ is a normalized effective environmental input mode.

The effective mode $\hat b_{\rm in}$ represents the particular normalized linear combination of the distributed environmental modes $\hat s(z,\omega)$ that couples to the detected waveguide mode. Together with the optical mode, it allows the lossy WGM channel to be represented as a unitary two-mode dilation. In particular, the input--output transformation may be completed as
\begin{equation}
\begin{pmatrix}
\hat a_5\\
\hat b_{\rm out}
\end{pmatrix}
=
\begin{pmatrix}
t(\delta) & \ell(\delta)\\
-\ell(\delta) & t^*(\delta)
\end{pmatrix}
\begin{pmatrix}
\hat a_4\\
\hat b_{\rm in}
\end{pmatrix},
\end{equation}
where $\ell(\delta)=\sqrt{1-|t(\delta)|^2}$. The first row is equivalent to Eq.~\eqref{eq:WGM_input_output}, with
$\hat F(\omega)=\ell(\delta)\hat b_{\rm in}(\omega)$, while the second row defines the corresponding effective environmental output mode $\hat b_{\rm out}$. Similarly, $\hat b_{\rm out}$ represents a normalised linear combination of environmental modes which photons in the WGM can couple to via loss mechanisms (physically, surface scattering and optical absorption in the WGM resonator). The transformation is unitary because $|t(\delta)|^2+\ell^2(\delta)=1$. Equivalently, inverting this transformation and taking the Hermitian conjugate gives the corresponding transformation of the input creation operator,
\begin{equation}
\hat a_{4}^{\dagger}\longrightarrow t(\delta)\hat a_{5}^{\dagger} -\ell(\delta)\hat b_{\mathrm{out}}^{\dagger},\qquad
\hat a_{3}^{\dagger}\longrightarrow\hat a_{6}^{\dagger}.
\label{wgm_output}
\end{equation}
The modes $\hat b_{\rm in}$ and $\hat b_{\rm out}$ are the input and output modes of the same effective loss channel. Since the environmental input is in the vacuum state at initial time, no photons enter the WGM from $\hat b_{\rm in}$ modes; nevertheless, this mode obeys the bosonic commutation relations. Correspondingly, since the WGM is occupied, photons circulating in the WGM can couple to the modes $\hat b_{\rm out}$. Thus, each photon lost from the WGM arm populates the output environmental mode $\hat b_{\rm out}$, while the reference arm (modes 3 and 6) is unchanged. Applying these transformations to the two-photon N00N state gives
\begin{equation}
\begin{aligned}
&|\psi_{5,6}\rangle=\frac{i}{\sqrt{2}}\left[t^{2}(\delta)|2,0\rangle_{5,6}+|0,2\rangle_{5,6}\right]\otimes|0\rangle_{\mathrm{env}}\\
&\quad -i\,t(\delta)\ell(\delta)|1,0\rangle_{5,6}\otimes|1\rangle_{\mathrm{env}}+\frac{i\ell^{2}(\delta)}{\sqrt{2}} |0,0\rangle_{5,6}\otimes|2\rangle_{\mathrm{env}} .
\end{aligned}
\label{eq:state_5_6}
\end{equation}
Here, $|1\rangle_{\mathrm{env}}$ and $|2\rangle_{\mathrm{env}}$ are normalized Fock states of the output environmental mode. Equation~\eqref{eq:state_5_6} is normalized, and since there is the doubled phase shift $2\arg[t(\delta)]$,  the two-photon amplitude in the WGM arm is multiplied by $t^{2}(\delta)$. Applying the second-beam splitter transformation to Eq.~\eqref{eq:state_5_6} gives the final output state in Eq.~\eqref{eq:two_photon_output_state}.

\section{Analytical Derivation of the Coherent- and Photon-Pair Quantum Fisher Information}
\label{app:qfi_derivation}

This appendix derives the QFI expressions quoted in Sec.~\ref{sec:general_qfi}. The optical output state and its loss-sector decomposition are derived in Sec.~\ref{sec:quantum_model} and Appendix~\ref{app:photon_pair_derivation}; here we begin directly from the reduced detected state. We use the polar decomposition introduced in Eq.~\eqref{eq:t_T_phi} and write $\dot T=\partial_\vartheta T$, $\dot\phi=\partial_\vartheta\phi$, and $\dot t=\partial_\vartheta t$. Because the second MZI beam splitter is independent of $\vartheta$, it does not affect the QFI.

\subsection{Photon-pair probe}
\label{app:qfi_pair}

Starting from the reduced state and loss-sector probabilities derived in Sec.~\ref{sec:quantum_model}, we write
\begin{equation}
\rho_{\mathrm{pair}} = \bigoplus_{n=0}^{2}p_n|u_n\rangle\langle u_n|,
\label{eq:app_pair_block_state}
\end{equation}
where $p_0=(1+T^2)/2$, $p_1=T(1-T)$, and $p_2=(1-T)^2/2$. A convenient choice of the normalized conditional optical states is
\begin{equation}
\begin{aligned}
|u_0\rangle&=\frac{T e^{2i\phi}|2,0\rangle+|0,2\rangle}{\sqrt{1+T^2}},\\
|u_1\rangle&=|1,0\rangle,\qquad
|u_2\rangle=|0,0\rangle.
\end{aligned}
\label{eq:app_conditional_states}
\end{equation}
The phase multiplying the one-photon-loss component is global within that sector and therefore disappears from $|u_1\rangle\langle u_1|$. Consequently, only the no-loss conditional state $|u_0\rangle$ carries nonzero conditional-state QFI.

For mutually orthogonal parameter-independent sectors, the QFI decomposes into the classical information contained in the sector probabilities and the average QFI within the conditional states:
\begin{equation}
F_Q(\rho_\vartheta)=\sum_n\frac{(\partial_\vartheta p_n)^2}{p_n}+\sum_n p_n
F_Q\!\left(|u_n\rangle\langle u_n|\right).
\label{eq:app_block_qfi}
\end{equation}
Applying Eq.~\eqref{eq:app_block_qfi} to Eq.~\eqref{eq:app_pair_block_state} therefore gives
\begin{equation}
F_Q^{\mathrm{pair}}=\sum_{n=0}^{2}\frac{\dot p_n^2}{p_n}+p_0F_Q(|u_0\rangle),
\label{eq:app_pair_qfi_decomposition}
\end{equation}
because $|u_1\rangle$ and $|u_2\rangle$ are parameter independent.

Using $\dot p_0=T\dot T$, $\dot p_1=(1-2T)\dot T$, and $\dot p_2=-(1-T)\dot T$, the contribution from the loss-sector probabilities reduces to
\begin{equation}
F_{\mathrm{prob}}\equiv \sum_{n=0}^{2}\frac{\dot p_n^2}{p_n}=\frac{\dot T^2}{T(1-T)}-\frac{2\dot T^2}{1+T^2}.
\label{eq:app_probability_qfi}
\end{equation}

For the no-loss state, writing $q=T e^{2i\phi}$ gives 
\begin{equation}
|u_0\rangle=\frac{1}{\sqrt{1+|q|^2}}(q|2,0\rangle+|0,2\rangle)    
\end{equation}
The pure-state QFI is then
\begin{equation}
F_Q(|u_0\rangle)=4 \frac{|\dot q|^2}{(1+|q|^2)^2}    
\end{equation}
Since $|q|^2=T^2$ and $|\dot q|^2=\dot T^2+4T^2\dot\phi^2$, its weighted contribution is
\begin{equation}
p_0F_Q(|u_0\rangle)=\frac{2\dot T^2}{1+T^2}+\frac{8T^2}{1+T^2}\dot\phi^2.
\label{eq:app_weighted_u0_qfi}
\end{equation}
Combining Eqs.~\eqref{eq:app_probability_qfi} and \eqref{eq:app_weighted_u0_qfi} in Eq.~\eqref{eq:app_pair_qfi_decomposition}, the terms proportional to $2\dot T^2/(1+T^2)$ cancel, yielding
\begin{equation}
F_Q^{\mathrm{pair}}(\vartheta)=\frac{\left(\partial_\vartheta T\right)^2}{T(1-T)}
+\frac{8T^2}{1+T^2}\left(\partial_\vartheta\phi\right)^2.
\label{eq:app_final_pair_qfi}
\end{equation}
This is the photon-pair QFI quoted in Sec.~\ref{sec:general_qfi}.

\subsection{Coherent-state probe}
\label{app:qfi_coherent}

A vacuum-loss channel maps a coherent input and a vacuum environmental state to a product of coherent states. Tracing over the environment therefore leaves a pure coherent state in the detected optical modes. For the multimode coherent state $|\boldsymbol{\zeta}(\vartheta)\rangle =\bigotimes_j|\zeta_j(\vartheta)\rangle$, the product of the single-mode coherent-state overlaps gives
\begin{equation}
\begin{aligned}
&\left|\left\langle\boldsymbol{\zeta}(\vartheta)\middle| \boldsymbol{\zeta}(\vartheta+d\vartheta) \right\rangle \right|^2=\exp\left[-\sum_j|d\zeta_j|^2\right] \\
&\qquad =1-\sum_j|\partial_\vartheta\zeta_j|^2d\vartheta^2+O(d\vartheta^3).
\end{aligned}
\end{equation}
For a differentiable family of pure states, the corresponding fidelity expansion is \cite{Braunstein_Caves1994}
\begin{equation}
\left|\langle\psi_\vartheta|\psi_{\vartheta+d\vartheta}\rangle \right|^2=
1-\frac{1}{4}F_Q(\vartheta)d\vartheta^2+O(d\vartheta^3).
\end{equation}
Comparing the second-order terms yields
\begin{equation}
F_Q(\vartheta)=4\sum_j|\partial_\vartheta\zeta_j|^2.
\label{eq:app_multimode_coherent_qfi}
\end{equation}
In the present WGM--MZI, the output amplitudes are 
\begin{equation}
\begin{aligned}
&\zeta_7=i\beta(1+t)/2 \\
&\zeta_8=\beta(1-t)/2.
\end{aligned}
\end{equation}
Hence, $\sum_{j=7,8}|\partial_\vartheta\zeta_j|^2 =|\beta|^2|\partial_\vartheta t|^2/2$, and the coherent-state QFI for arbitrary $\beta$ is $F_Q^{\mathrm{coh}}=2|\beta|^2|\partial_\vartheta t|^2$.

Under the equal-resource condition $|\beta|^2=2$, and using the polar decomposition in Eq.~\eqref{eq:t_T_phi}, this becomes
\begin{equation}
F_Q^{\mathrm{coh}}(\vartheta)=4|\partial_\vartheta t|^2= \frac{\left(\partial_\vartheta T\right)^2}{T}+4T\left(\partial_\vartheta\phi\right)^2.
\label{eq:app_final_coherent_qfi}
\end{equation}
This is the coherent-state QFI quoted in Sec.~\ref{sec:general_qfi}.

\subsection{Boundary points}
\label{app:qfi_limiting_cases}

Equations~\eqref{eq:app_final_pair_qfi} and \eqref{eq:app_final_coherent_qfi} apply at regular points with $0<T<1$. At a rank-changing boundary, the exact SLD QFI must instead be evaluated directly from the physical density operator and its derivative. In particular, at $r=\alpha$ and $\delta=0$, one has $\partial_\alpha\rho_{\mathrm{pair}}=0$, so the exact photon-pair SLD QFI for estimating $\alpha$ is zero. The values of QFI calculated at nearby regular points approach a finite limit as previously unoccupied loss sectors acquire probabilities at second order in the parameter displacement. This limiting value describes the QFI at nearby regular points and should not be identified with the QFI at the exact boundary point. 
At $T=1$, the loss probabilities vanish, so the QFI must be evaluated directly at the boundary or defined through a clearly specified limit.

\section{WGM Linewidth and Q Factor}
\label{app:wgm_linewidth}

In WGM sensing experiments, it is more typical to characterise a WGM resonator and coupling conditions using its intensity linewidth (FWHM) in wavelength units $\Delta \lambda_{\mathrm{FWHM}}$ and its Q factor. In the narrow-resonance approximation with constant refractive index $n$, the dimensionless parameters $\alpha, r$ can be related to these quantities through the following expressions:
\begin{equation}
    \Delta \lambda_{\mathrm{FWHM}} \simeq - \dfrac{\lambda_o^2 \ln(|\alpha r|)}{2\pi^2 n R} ,
\end{equation}
and
\begin{equation}
    Q = \dfrac{\lambda_o}{\Delta \lambda_{\mathrm{FWHM}}} \simeq - \dfrac{2\pi^2 n R}{\lambda_o \ln(|\alpha r|)} .
\end{equation}
The WGM resonator radius is $R$ and the resonance wavelength $\lambda_o$.

\end{document}